\documentclass[twocolumn,pre,floatfix]{revtex4}
\usepackage{epsfig}
\usepackage{amsfonts}
\usepackage{amssymb}
\usepackage{amsmath}
\usepackage{yfonts}
\usepackage{pgf,pgfarrows,pgfnodes,pgfautomata,pgfheaps,pgfshade}
\usepackage{graphics,graphicx}

\newcommand{\Z}{{\mathbb{Z}}}

\begin{document}
\title{Compact stationary fluxons in the Josephson junction ladder}
\author{Andrii O. Prykhodko}
\affiliation{Kyiv Academic University, 36 Vernadsky blvd., Kyiv 03142, Ukraine}
\affiliation{Bogolyubov Institute for Theoretical Physics,
National Academy of Sciences of Ukraine,
vul. Metrologichna 14B, 
 03143 Kyiv, Ukraine}
\author{Ivan O. Starodub}
\affiliation{Bogolyubov Institute for Theoretical Physics,
National Academy of Sciences of Ukraine,
vul. Metrologichna 14B, 
 03143 Kyiv, Ukraine} 
\author{Yaroslav Zolotaryuk}
\email{yzolo@bitp.kyiv.ua}
\affiliation{Bogolyubov Institute for Theoretical Physics,
National Academy of Sciences of Ukraine,
vul. Metrologichna 14B, 
 03143 Kyiv, Ukraine}
\date{\today}

\begin{abstract}
Stationary  compact fluxon profiles are shown to be exact
solutions of the inductively coupled and dc-biased 
Josephson junction ladder.  Such states do not exist 
in the parallel Josephson junction array which is described by the
standard discrete sine-Gordon equation. 
It is shown that there are compact fluxon and 
multi-fluxon states which either satisfy the top-bottom antisymmetry
or are asymmetric. The anti-symmetric states
have zero energy if their topological charge is even and the
asymmetric states always have zero energy. Depending on the
anisotropy constant the compact fluxons can either coexist with the
non-compact states or only compact states are possible.
External magnetic field prevents compact state existence.

\end{abstract}
\maketitle

\section{Introduction} \label{intro}

Josephson junction ladders (JJLs) have attracted a significant interest
from researches because of their non-trivial statistical
\cite{k84prb,dt95prl} and dynamical properties \cite{fmamfm98pd}. 
There are several non-trivial wave phenomena that exist in JJLs and
are not possible in the standard parallel arrays which a described
by the discrete sine-Gordon (or Frenkel-Kontorova) model \cite{wzso96pd}.
Discrete breathers or intrinsic localized modes
were discovered experimentally \cite{tmo00prl,baufz00prl} and this
discovery opened a new chapter in the studies of the Josephson
junction based systems. There is a flat band in the JJL
plasmon spectrum \cite{mffzp01pre,bz22jpc}, and, again, no such
a thing is possible in the parallel array. 
Starting from the pioneering work \cite{mff95prbII} a significant number
of results have been obtained concerning the stationary 
vortex (or fluxon) properties in JJLs under the influence of the
external magnetic field. Stationary fluxon properties in JJL were
investigated in \cite{gfpg96pla}. Trias and coworkers \cite{tmo02prb} 
performed detailed studies of the  fluxon-discrete breather
interactions. 

In this work we demonstrate that an important family of stationary
fluxon states in JJLs was overlooked. Since an JJL has a flat
band in its linear spectrum, it is logical to expect it to have,
alongside with the exponentially decaying fluxons, completely
compact fluxons. Under the term "compact" we understand the exact
step-like solutions which has the phase distribution of the
type $0,\cdots, 0,2\pi, 2\pi, \cdots, 2\pi$. We will demonstrate
the existence of these solutions, compare them to the non-compact
fluxons and compute the energy of the both. It will be shown that their
existence depends strongly on the JJL anisotropy.

The importance of this work can be viewed from another angle. Researchers
working in the area of nonlinear lattice dynamics study
discrete compactons \cite{kkbt02jpa,aks10prl}, solutions that are localized
on several particles while the rest of the lattice is at equilibrium. 
The solutions we are discussing in this paper are not dynamical, 
they are static. However, they have topological charge. 
To our knowledge, this is the first report about
compact topological excitations in the realistic physical system.

\section{Model and equations of motion} \label{model}

\begin{figure}[h]
\setlength{\unitlength}{9.5cm}
\begin{picture}(1,0.6)
\thicklines
\put(0.05, 0.10){\line(1, 0){0.45}}
\put(0.05, 0.3){\line(1, 0){0.45}}
\put(0.55, 0.1){\makebox(0,0){$\Huge \cdots$}}
\put(0.55, 0.3){\makebox(0,0){$\Huge \cdots$}}
\multiput(0.6, 0.1)(0,0.2){2}{\line(1, 0){0.25}}

\multiput(0.05,0.1)(0.2,0){3}{\line(0, 1){0.30}}
\put(0.65,0.10){\line(0,1){0.3}}
\put(0.85,0.10){\line(0,1){0.3}}
\multiput(0.05,0.1)(0.2,0){3}{\vector(0,-1){0.13}}
\multiput(0.65,0.1)(0.2,0){2}{\vector(0,-1){0.13}}
\multiput(0.05,0.43)(0.2,0){5}{\vector(0,-1){0.13}}
\multiput(0.02, 0.02)(0.2,0){3}{\makebox(0,0){$\Large I_B$}}
\multiput(0.62, 0.02)(0.2,0){2}{\makebox(0,0){$\Large I_B$}}
\multiput(0.02, 0.4)(0.2,0){3}{\makebox(0,0){$\Large I_B$}}
\multiput(0.62, 0.4)(0.2,0){2}{\makebox(0,0){$\Large I_B$}}

\multiput(0.15,0.10)(0.2,0){2}{\makebox(0,0){$\Huge \times$}}
\multiput(0.15,0.30)(0.2,0){2}{\makebox(0,0){$\Huge \times$}}
\multiput(0.75,0.30)(0.,-0.2){2}{\makebox(0,0){$\Huge \times$}}
\multiput(0.05,0.20)(0.2,0){5}{\makebox(0,0){$\Huge \times$}}

\put(0.15, 0.06){\makebox(0,0){$\large \phi^{(h)}_{1,1}$}}
\put(0.15, 0.34){\makebox(0,0){$\large \phi^{(h)}_{1,2}$}}
\put(0.1, 0.2){\makebox(0,0){$\large \phi^{(v)}_{1}$}}

\put(0.35, 0.06){\makebox(0,0){$\large \phi^{(h)}_{2,1}$}}
\put(0.35, 0.34){\makebox(0,0){$\large \phi^{(h)}_{2,2}$}}
\put(0.3, 0.2){\makebox(0,0){$\large \phi^{(v)}_2$}}

\put(0.75, 0.06){\makebox(0,0){$\large \phi^{(h)}_{N,1}$}}
\put(0.75, 0.34){\makebox(0,0){$\large \phi^{(h)}_{N,2}$}}
\put(0.7, 0.2){\makebox(0,0){$\large \phi^{(v)}_N$}}
\put(0.9, 0.2){\makebox(0,0){$\large \phi^{(v)}_{N+1}$}}

\end{picture}
\caption{Schematic view of the dc-biased
2-row ladder with open ends. Crosses correspond to the Josephson
junctions.}
\label{fig1}
\end{figure}
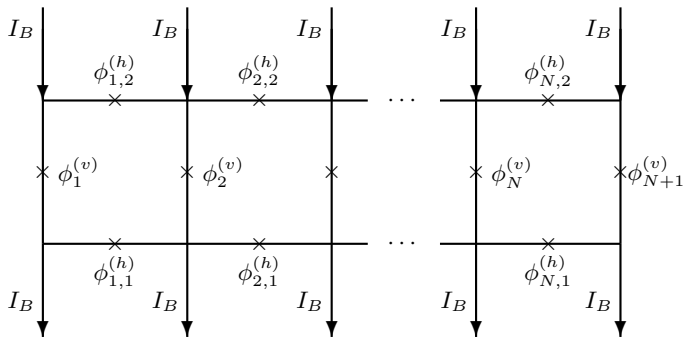

In this work the 2-row ladder of small Josephson junctions is studied.
The elementary cell of the
 Josephson junction ladder consists of three junctions,
 as shown in Fig. \ref{fig1}. Dynamics of each of these junctions is described by the
  Josephson phase which is the difference
 of the superconducting wavefunction phases. The Josephson phases
that belong to the $n$th elementary cell will be denoted as follows:
the vertical phase 
$\phi^{(v)}_n$, the bottom horizontal  $\phi^{(h)}_{n,1}$ and the 
top  horizontal $\phi^{(h)}_{n,2}$ phases. In total the ladder consists
of $N+1$ vertical junctions, $N$ top horizontal and $N$ bottom 
horizontal junctions. Each vertical junction is biased by the external
dc current, $I_B$. 
%
%
%

The Hamiltonian function of the whole JJL is based on the
resistively and capacitively shunted model and can be written as follows:
\begin{eqnarray}
\!\!\!\!\!\!\!\!\!\!\!\!\! \label{en1}
&&H=E_K+U_J+U_L,\\
\!\!\!\!\!\!\!\!\!\!\!\!\! \label{en2}
&& E_K= \frac{C_v}{2} \sum_{n=1}^{N+1} \left(
\frac{\hbar \dot{\phi}^{(v)}_n}{2e} \right)^2+ \frac{C_h}{2}  \sum_{n=1}^{N} \sum_{k=1}^{2}\left(
\frac{\hbar \dot{\phi}^{(h)}_{n,k}}{2e}\right)^2,\\
\!\!\!\!\!\!\!\!\!\!\!\!\!
\nonumber
&& U_J=\sum_{n=1}^{N+1} \left[ E^{(v)}_J (1-\cos{\phi}^{(v)}_n )-I_B{\phi}^{(v)}_n \right]+   \\
\!\!\!\!\!\!\!\!\!\!\!\!\! \label{en3}
&&+\sum_{n=1}^{N}\sum_{k=1}^2 E^{(h)}_J (1-\cos{\phi}^{(h)}_{n,k} ),\\
\!\!\!\!\!\!\!\!\!\!\!\!\! 
&& U_L=\frac{\Phi^2_0}{8\pi^2L}\sum_{n=1}^{N}  
\left( \phi^{(v)}_{n+1}-\phi^{(v)}_{n}+\phi^{(h)}_{n,2}-\phi^{(h)}_{n,1}-2\pi f\right)^2.\label{en4}
\end{eqnarray}
Here $E^{(v,h)}_J=\hbar I^{(v,h)}_c/(2e)$ is the Josephson coupling
energy for the vertical (superscript $^{(v)}$) or horizontal 
(superscript $^{(h)}$) junctions. 
The  critical currents and capacitances of the junctions
are given by $I^{(v,h)}$ and $C_{v,h}$, respectively. The frustration
parameter $f=\Phi_e/\Phi_0$ is the ratio of the external magnetic
field flux to the flux quantum $\Phi_0=\pi\hbar/e$. 
As one can see, the total energy is divided into three parts: the 
total energy all of the capacitors, $E_K$, the total Josephson
energy, $U_J$, and the inductive energy, stored in all JJL cells (loops), 
$U_L$. The JJL can be interpreted as a $6N+2$-dimensional 
hamiltoninan system with
$E_K$ being its kinetic energy and $U_J+U_L$ being its potential energy.

The equations of motion for the Josephson phases of the ladder can either be deduced from the above Hamiltonian or can be obtained
from the Kirhoff equations, Josephson equations for each of the juncions
and the flux quantization realtion for each loop.
Since these equations of motion have been derived in the number of previous papers \cite{baufz00prl,tmo02prb} we will not repeat the derivation
procedure here. 
We introduce the dimensionless time $t\to t\omega_p$, where  
$\omega_p =\sqrt{{2 \pi I_c^{(v)}}/{(\Phi_0 C_v)}}$ is the
Josephson plasma frequency. We also introduce the following dimensionless
parameters:
\begin{eqnarray}
\nonumber
&&\frac{1}{\beta_L}= \frac{\Phi_0}{2\pi L I^{(v)}_c},\;
\gamma=\frac{I_B}{I^{(v)}_c},\; \eta=\frac{I^{(h)}_c}{I^{(v)}_c}=
\frac{C_h}{C_v}=\frac{R_v}{R_h}, \\
&& \label{dim}
\;\alpha=\frac{\hbar\omega_p}{2e R_v I^{(v)}},
\end{eqnarray}
where $R_{v,h}$ is the vertical (horizontal) junction resistance, $\beta^{-1}_L$ describes the JJL discreteness, $\gamma$ is dimensionless
external current, $\eta$ is the anisotropy constant and $\alpha$ is the
dissipation parameter associated the normal electron tunneling.  
The anisotropy constant describes whether the energy 
is concentrated mostly 
in the vertical subsystem ($\eta \ll1 $) or in the horizontal one ($\eta \gg 1$). Finally, 
one ends up with the dimensionless set of evolution equations
which are, in fact, a system of coupled discrete sine-Gordon
(DSG) equations. For the sake of brevity we introduce 
\begin{equation}
 {\cal N} (x) \equiv \ddot{x} + \alpha \dot{x} + \sin x .
\end{equation}
The final equations have the following form:
\begin{eqnarray}\label{emd1}
&& {\cal N} (\phi^{(v)}_{n}) =\gamma+\frac{1}{\beta_L} \left (\hat{\mathfrak{D}} \phi^{(v)}_n+ \right .\\
\nonumber
&& \left. +\phi^{(h)}_{n,2}-\phi^{(h)}_{n,1}-\phi^{(h)}_{n-1,2}+\phi^{(h)}_{n-1,1}\right ), \,n=\overline{1,N+1},\\
\label{emd2}
&& {\cal N} ({\phi}^{(h)}_{n,k}) = -
(-1)^k\frac{\Delta_{n+1,n}}{\eta\beta_L} ,\; k=1,2,\\
\label{emd3}
&& \Delta_{n+1,n} \equiv
\phi^{(v)}_{n+1}-\phi^{(v)}_{n} + \phi^{(h)}_{n,2}- \phi^{(h)}_{n,1}-2\pi f ,  \label{emd5} 
\end{eqnarray}
together with the boundary conditions:
\begin{eqnarray}\label{emd6}
&&{\cal N}(\phi^{(v)}_{1})=\gamma+\frac{1}{\beta_L} \Delta_{2,1} ,\\
\label{emd7}
&&{\cal N}({\phi}^{(v)}_{N+1})=\gamma-\frac{1}{\beta_L} \Delta_{N+1,N},
\end{eqnarray}
where $n=\overline{1,N}$ stands for $n=1,2,\ldots, N$.
The discrete laplacian operator in Eq. (\ref{emd1}) is defined
as $\hat{\mathfrak{D}} \phi_n \equiv \phi_{n+1}-2\phi_n+\phi_{n-1}$.
A careful look at Eq. (\ref{emd2}) reveals that the dynamical
equations for the top and bottom junctions reproduce each other
when the top-bottom inversion transform 
$\phi^{(h)}_{n,1} \to -\phi^{(h)}_{n,2}$ is applied. However, we are
not going  to exclude one of the horizontal
components with the help of this antisymmetry and will solve the full system of equations.

The dimensionless JJL energy is obtained by introducing the parameters
(\ref{dim}) and normalizing the energy (\ref{en1})-(\ref{en4}) to the
Josephson energy of the vertical junction ($E^{(v)}_J$):
\begin{eqnarray}\label{ham2}
&&H=\sum_{n=1}^N\frac{1}{2} \left [{\dot{\phi}_n^{(v)}}{}^2 +
\eta \sum_{k=1}^2 {\dot{\phi}_{n,k}^{(h)}}{}^2 \right ]+
\frac{1}{2} {\dot{\phi}_{N+1}^{(v)}}{}^2+\\
\nonumber
&&+\frac{1}{2\beta_L}\sum_{n=1}^{N+1}\left( \phi^{(v)}_{n+1}-\phi^{(v)}_{n}+
\phi^{(h)}_{n,2}-\phi^{(h)}_{n,1}-2\pi f\right)^2 +\\
\nonumber
&&+ \sum_{n=1}^{N+1} \left[1-\cos{\phi}^{(v)}_n -\gamma{\phi}^{(v)}_n \right]+ 
\eta \sum_{n=1}^{N}\sum_{k=1}^2 (1-\cos{\phi}^{(h)}_{n,k} ).
\nonumber
\end{eqnarray}
From this formula one can clearly understand  the physical meaning
of the anisotropy constant $\eta$. It characterizes the 
ratio of the energy stored in the subsystem of horizontal junctions
to the energy of the vertical subsystem.

\section{Properties of compact and non-compact fluxons} \label{vort}

In this section the properties of the compact states as well as their
differences with respect to the non-compact fluxons.

\subsection{Compact fluxons}

Fluxons are topological excitaions that connect two different ground states 
of the JJL. They carry the integer number of the magnetic flux quanta and are characterized by the topological charge, $Q \in \Z$. For example, the fluxon with $Q=1$
connects the minima with $\phi^{(v)}_n=\arcsin \gamma$ and
$\phi^{(v)}_n=2\pi +\arcsin \gamma$. 
Careful analysis of the r.h.s. of Eqs. (\ref{emd1})-(\ref{emd7}) shows
that compact solutions can be exist as fixed points of the system if no 
magnetic field is applied ($f=0$). 
Consider the stationary case $\dot{\phi}^{(v)}_n=0$ ($n=\overline{1,N+1}$), 
$\dot{\phi}^{(h)}_{n,k}=0$ ($n=\overline{1,N1}$, $k=1,2$).
If all the vertical phases correspond to the solutions of the
equation $\sin \phi^{(v)}_n=\gamma$ and the horizontal phases
satisfy $\sin \phi^{(h)}_{n,m}=0$, $m=1,2$, the conditions for the
stationary configuration reduces to 
\begin{eqnarray}
\nonumber
&&\hat{\mathfrak{D}} \phi^{(v)}_n\equiv 
(\phi_{n+1}^{(v)}-\phi_{n}^{(v)})-(\phi_{n}^{(v)}-\phi_{n-1}^{(v)})=\\
\label{17}
&&= (\phi^{(h)}_{n,1}-\phi^{(h)}_{n-1,1})-(\phi^{(h)}_{n,2}-\phi^{(h)}_{n-1,2}), \\
\label{18}
&& \phi^{(v)}_{n+1}-\phi^{(v)}_{n} =  \phi^{(h)}_{n,1}- \phi^{(h)}_{n,2}.
\end{eqnarray} 
One can easily observe that Eq. (\ref{17}) can be obtained
by subtracting Eq. (\ref{18}) for $n$ and $n+1$.
At this point we notice that there are two possibilities
to have stationary compact fluxon solutions:
\begin{itemize}
\item[(i)] If one takes into account the fact that $\phi^{(h)}_{1,n}=-\phi^{(h)}_{2,n}$, 
which follows directly from Eqs. (\ref{emd1})-(\ref{emd7}), 
we arrive to the single equation
\begin{equation}\label{19}
\phi^{(h)}_{n,1}=\frac{1}{2} \left (\phi^{(v)}_{n+1}-\phi^{(v)}_{n} \right),
n=\overline{1,N} .
\end{equation}
The states that satisfy Eq. (\ref{19}) will be referred to as
anti-symmetric because they are anti-symmetric with respect to the
"top"-"bottom" inversion.
\item[(ii)] There is another possibility. One can set either 
$\phi^{(h)}_{n,1}=0$, or $\phi^{(h)}_{n,2}=0$  for   $n=\overline{1,N}$.
Without loss of generality assume  $\phi^{(h)}_{n,2}=0$, then
\begin{equation}\label{20}
\phi^{(h)}_{n,1}=\phi^{(v)}_{n+1}-\phi^{(v)}_{n}, n=\overline{1,N}. 
\end{equation}
\end{itemize}
The conditions (\ref{19}) or (\ref{20}) 
can produce the infinite number of stationary states.
In this work we will restrict ourselves to the elementary topological
states (fluxons) and their most trivial combinations.
\paragraph*{Elementary fluxons.}
These are the step-like solutions that connect two different ground states
at the left and right sides of the ladder. 
The anti-symmetric fluxons can be written explicitly as
\begin{eqnarray}\!\!\!\!\!\!\!\!\!\!\!\!\!
&&\phi^{(v)}_n=\left\{ \begin{array}{cc}
\arcsin \gamma,& n\le n_0,\\
2\pi Q+\arcsin\gamma,& n > n_0 
\end{array} \right., \,n=\overline{1,N+1}, \label{vert1}\\
\!\!\!\!\!\!\!\!\!\!\!\!\!\!
&&\phi^{(h)}_{n,1}=-\phi^{(h)}_{n,2}=\pi Q \delta_{nn_0},\;\; n=\overline{1,N}.
\label{antsym1}
\end{eqnarray}
The asymmetric compact states have the same distribution of the vertical phases
but the horizontal phases must satisfy 
\begin{eqnarray}
&&\phi^{(h)}_{n,1}=2\pi Q \delta_{nn_0},\; \phi^{(h)}_{n,2}=0,\; n=\overline{1,N},
\label{asym1}\\
&& \mbox{or} \nonumber \\
&&\phi^{(h)}_{n,1}=0,\; \phi^{(h)}_{n,2}=-2\pi Q \delta_{nn_0},\;  n=\overline{1,N}
\label{asym2}
\end{eqnarray}
These asymmetric states serve as a good demonstration of the importance
of the full system of equations of motion that include both the top
and bottom subsystems (dropped in Ref. \cite{gfpg96pla}). 
Despite the top-bottom antisymmetry of
the equations of the motion (\ref{emd1})-(\ref{emd7}), there are dynamical solutions 
like discrete breathers \cite{baufz00prl,bau00pre,tmo02prb} 
that break this symmetry.
Since the values of the horizontal and vertical phases are connected,
we can define a coding sequence that uses the vertical phase distribution
in order to describe the total solution.
For example, the $4\pi$ elementary fluxon will be denoted as $(\cdots 0, 4\pi, \cdots)$
and the $2 \pi$ antifluxon will be denoted as $(\cdots 0, -2\pi, \cdots)$.

We 
reiterate that these compact states are exact solutions of the equations of motion (\ref{emd1})-(\ref{emd7}). They are not possible for the
standard parallel Josephson junction array that is
governed by the DSG equation ${\cal N}(\phi_n)=\mathfrak{D}\phi_n+\gamma$.
In DSG it is not possible to have a compact state, say $(\cdots 0,2\pi,\cdots)$,
because the phases decay asymptotically to the values $0$ and $2\pi$ 
as $n\to \pm \infty$. The inductive energy $\propto \sum_n (\phi_{n+1}-\phi_n)^2$ can not reach its global minimum, however, in the JJL
case the inductive energy is $\propto (\phi^{(v)}_{n+1}-\phi^{(v)}_{n}+
\phi^{(h)}_{n,2}-\phi^{(h)}_{n,1})^2$. From here one can see that it is possible 
to nullify the energy by choosing the horizontal phases appropriately.
 
\paragraph*{Combinations of the elementary fluxons.}
It is straightforward to conclude, that all possible superpositions
of compact fluxons and antifluxons separated by an arbitrary number
of sites will also be the stationary solutions
of the equations of motion. The $M$-fluxon or a multi-fluxon 
state that
consists of $M$ fluxons with charges $Q_j>0$, $j=\overline{1,M}$, centered 
at the points $1<n_1<n_2 <\cdots<n_M<N$ has the following distribution
of the vertical phases:
\begin{eqnarray}\!\!\!\!\!\!\!\!\!
&&\phi^{(v)}_n=\left\{ \begin{array}{cc}
\arcsin \gamma,& n\le n_1,\\
2\pi Q_1+\arcsin\gamma,&  n_1  <n \le n_2,\\
2\pi (Q_1+Q_2)+\arcsin\gamma,&  n_2  <n \le n_3,\\
\cdots \\
2\pi \sum_{m=1}^MQ_j+\arcsin\gamma,&  n> n_{M},\\
\end{array} \right. \label{vert2} \\
\!\!\!\!\!\!\!\!\!
&& \,n=\overline{1,N+1}. \nonumber
\end{eqnarray}
And the distribution of the 
horizontal phases, for both the anti-symmetric and asymmetric fluxons, 
are given by the folowin expression
\begin{eqnarray}
\mbox{Anti-sym.:}&&\phi^{(h)}_{n,1}= \pi \sum_{j=1}^M Q_j \delta_{n,n_j}= -\phi^{(h)}_{n,2}, \label{hor2a}\\
\mbox{Asym.:}&& \phi^{(h)}_{n,1}= 2\pi \sum_{j=1}^M Q_j \delta_{n,n_j}, \phi^{(h)}_{n,2}=0. \label{hor2b}
\end{eqnarray}
It should be noted that the positions of the fluxons that constitute a
combiden state should not coincide.
Suppose we have $n_1=n_2$, then we will obtain an elementary fluxon (\ref{vert1}) with 
the total charge $Q_1+Q_2$. Construction of the $M$-antifluxon 
state with topological  charges $Q_j<0$,$j=\overline{1,M}$ is 
straightforward.

It is possible to construct states with total zero topological charge. For
example, the bound state of two oppositely charged fluxons with $\nu>0$
vertical junctions between them is expressed as 
\begin{eqnarray}
\!\!\!\! \label{25}
 \phi^{(v)}_n= \left\{ \begin{array}{cc} 
\arcsin \gamma,& n\le n_0, n>n_0+\nu,\\
2\pi Q+\arcsin\gamma,&  n_0  <n \le n_0+\nu,\\
\end{array}\right.  \\ 
\!\!\!\! n=\overline{1,N+1}, \nonumber \\
\!\!\!\! \label{26}
\mbox{Antisym.:}\;\;  \left. \begin{array}{c}
\phi^{(h)}_{n,1}= \pi Q \delta_{n,n_0}=-\phi^{(h)}_{n_0,2}, \\
\phi^{(h)}_{n_0+\nu,1}= -\pi Q \delta_{n,n_0+\nu}=-\phi^{(h)}_{n_0+\nu,2},
\end{array} \right.
\\
\!\!\!\! \label{27}
\mbox{Asym.:} \;\;\phi^{(h)}_{n,1}= 2\pi Q \delta_{n,n_0}=-\phi^{(h)}_{n_0+\nu,2};\;\; \phi^{(h)}_{n,2}= 0, \\
\!\!\!\!   n=\overline{1,N}\nonumber .
\end{eqnarray}
One can move further by creating various 
combinations of states (\ref{vert2})-(\ref{hor2b}) and (\ref{25})-(\ref{27}).

The next logical step is to investigate the stability of these compact
solutions. There are two ways of doing that.
One way is to minimize numerically the ladder energy
(\ref{ham2}) and find the stable configurations. We have used another 
approach (also used in Ref. \cite{gfpg96pla}) which was
to solve numerically the equations of motion until the system settles
on the stable stationary solution (attractor). 
The equations of motion (\ref{emd1})-(\ref{emd7}) were solved numerically 
using the 4th order Runge-Kutta method with the 
initial condition that corresponds to the non-compact fluxon
in the vertical subsystem and unperturbed horizontal subsystem:
\begin{eqnarray}
&&\phi^{(v)}_n=\arcsin \gamma + 4Q \arctan e^{\frac{n-n_0+\delta}{d}},\;
n=\overline{1,N+1}, \nonumber \\
&&\phi^{(h)}_{n,m}=0,\; m=1,2, \; n=\overline{1,N}. \label{ic1}
\end{eqnarray}
Here $d$ is initial fluxon width, $n_0$ is its initial position and
$0\le \delta \le 1/2$ is used to produce a junction-centered ($\delta=0$) or
between-junction-centered or cell-centered ($\delta=1/2$) stationary fluxon.
The initial width is chosen according to the continuum sine-Gordon
theory $d=1/\sqrt{\beta_L}$. In these simulations
the dissipation parameter $\alpha$  controls only the speed of 
relaxation to the equilibrium, and, therefore, can be chosen arbitrarily.
We take it $\alpha=0.25$ unless stated otherwise.

\begin{figure}[ht]
\includegraphics[scale=0.27]{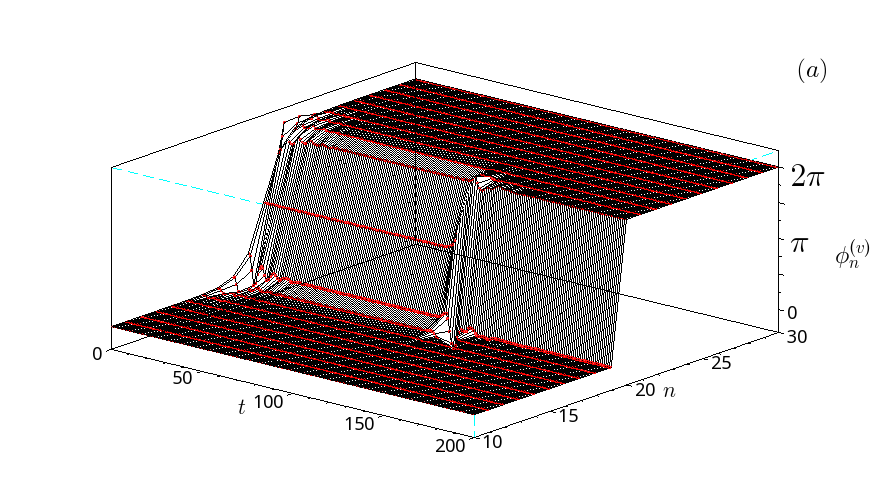}
\includegraphics[scale=0.27]{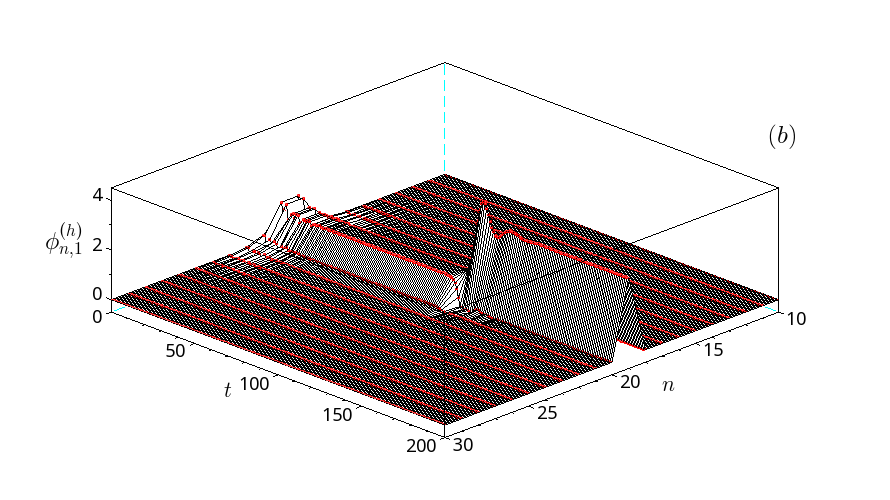}
\caption{Fluxon profile compactification for the vertical (a) and bottom
horizontal (b) junctions as a function of time
for the discreteness $\beta_L=1.5$, bias $\gamma=0$ and anisotropy $\eta=0.5$. The ladder size is $N=40$.}
\label{fig2}
\end{figure}

The results of such simulations are presented in Fig. \ref{fig2}.
%
%
What is observed can be called the {\it compactification} process when
the initially non-compact profile gradually becomes more localized
and, finally, ends up as an anti-symmetric compact fluxon that satisfies
Eqs. (\ref{vert1})-(\ref{antsym1}). The time evolution of the top
horizontal junctions was not plotted for the sake of brevity.

\begin{figure}[ht]
\includegraphics[scale=0.33]{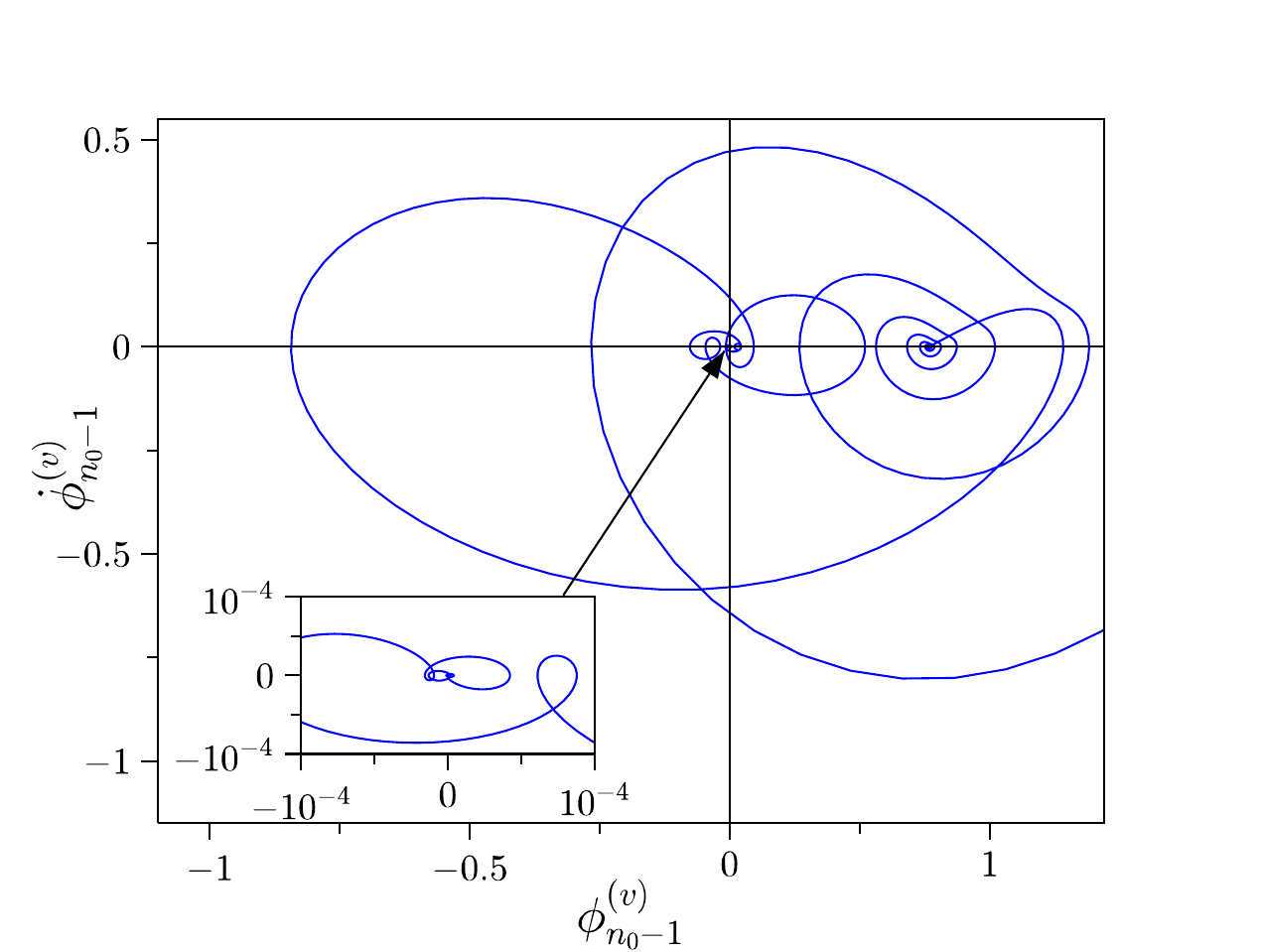}
\caption{Time evolution of the $(n_0-1)$ vertical junction for the
case described in Fig. \ref{fig2}.}
\label{fig3}
\end{figure}

 A detailed analysis of the vertical phase behavior during the
 compactification process is given in Fig. \ref{fig3}. The time
evolution of the $(n_0-1)$th junction on the 
$(\phi^{(v)}_{n_0-1},\dot{\phi}^{(v)}_{n_0-1})$ phase plane shows that
it indeed converges to the zero value.
%
%

\begin{figure}[ht]
\includegraphics[scale=0.27]{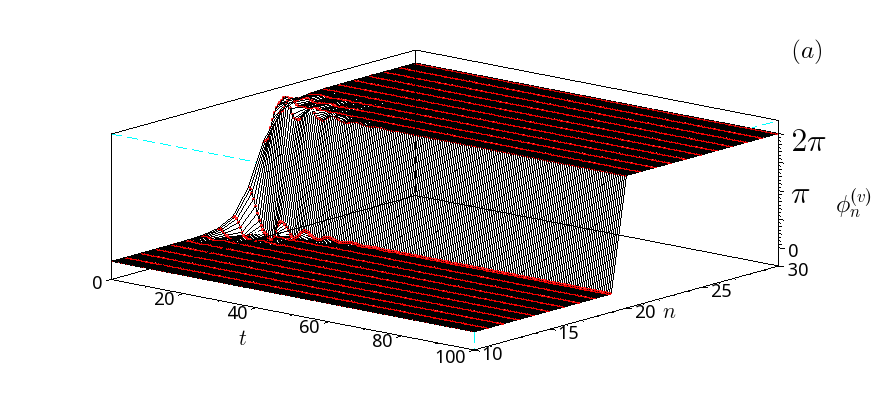}
\includegraphics[scale=0.27]{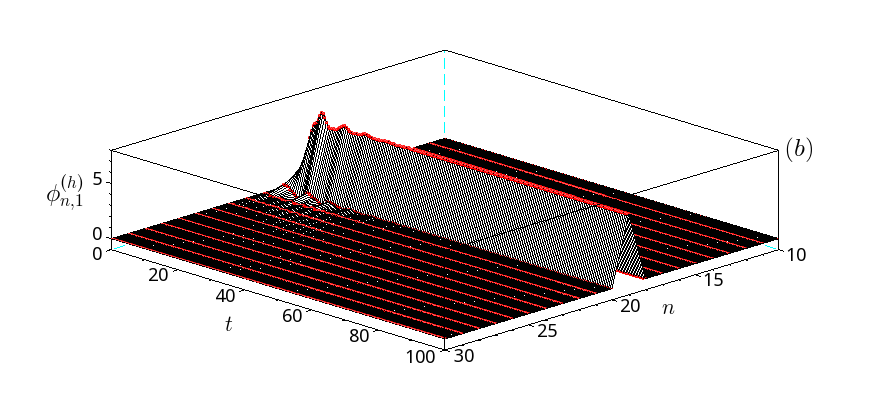}
\includegraphics[scale=0.27]{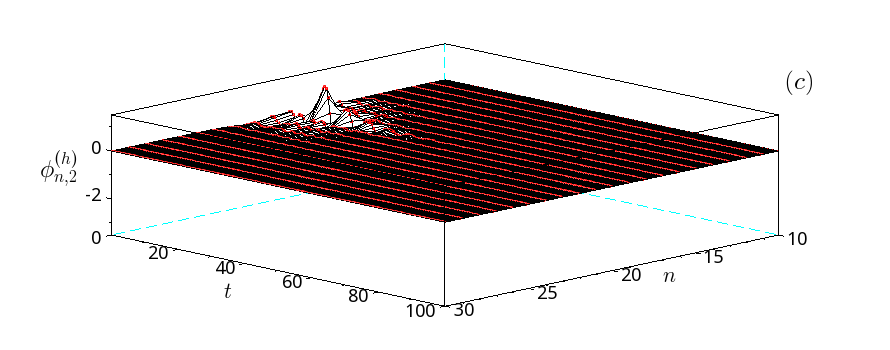}
\caption{Fluxon profile compactification for the vertical (a), 
horizontal bottom (b) and top (c) junctions as a function of time
for the same parameters as in Fig. \ref{fig2}.}
\label{fig4}
\end{figure}

It is possible to observe the fluxon compactification to the 
asymmetric state given by Eqs. (\ref{vert1}), (\ref{asym1})-(\ref{asym2}). 
In this case the initial conditions were modified. The initial vertical phase 
distribution was chosen as in Eq. (\ref{ic1}), the bottom horizontal
was phases were chosen as
\begin{eqnarray}\label{ic2}
\phi^{(h)}_{n,1}=0,\;  \phi^{(h)}_{n,2}=-\frac{1}{\cosh \left( \frac{n-n_0+\delta}{d}\right)}, \;n=\overline{1,N}.
\end{eqnarray}
For the simulations we have taken $\delta=0.5$. The results are
shown in Fig. \ref{fig4}. One can observe that the compatification process
is happening rather fast and the final configuration becomes compact and
 asymmetric (with respect to the top-bottom inversion) as prescribed by Eqs. (\ref{vert1}), (\ref{asym1})-(\ref{asym2}).
%
%
It is seen very clearly that the central bottom horizontal junction
tends to the value $2\pi$ while the rest approach $0$. The top 
horizontal subsystem junctions tends to the unexcited ground
state $\phi^{(h)}_{n,2}=0$. 

We have performed similar simulations for the different values of 
$\eta$ and $\beta_L$ as well as for the higher topological charges.
We have observed that the compactification process happens for
smaller values of $\eta$ while for larger anisotropies the system
ends on the non-compact state. We find it to be useful to investigate the
dependence of the compactification process on the anisotropy constant
$\eta$
together with calculating the stationary fluxon energy.

\subsection{Energies of the compact and non-compact fluxons}

The energies of the compact stationary fluxons can be calculated 
analytically by substituting the exact solutions (\ref{vert1}),
(\ref{antsym1}) and (\ref{asym1}-\ref{asym2}) into the JJL energy
(\ref{ham2}). 
The energies of the elementary anti-symmetric ($E_{ANS}$) and 
asymmetric ($E_{A-S}$) fluxons for a given topological charge $Q\in \Z$ 
can be written as
\begin{eqnarray}
\label{enq1}
&&E_{ANS}[Q]= 2[1-(-1)^{Q}] \eta+D_Q(\gamma,n_0,N),\\
\label{enq2}
&&E_{A-S}[Q]=D_Q(\gamma,n_0,N),\\
\nonumber
&&D_Q(\gamma,n_0,N)=N(1-\sqrt{1-\gamma^2}-\gamma\,\arcsin\gamma)-\\
&& -2\pi Q(N-n_0)\gamma.\label{D}
\end{eqnarray}
Here the term $D_Q(\gamma,n_0,N)$ appears because the dc bias
makes the potential minima for the $\phi=\arcsin \gamma$
and $\phi=2\pi Q +\arcsin \gamma$ non-equivalent. Naturally,
$D_Q(\gamma=0,n_0,N)=0$. Let us assume for a while that
the $\gamma=0$. The energy for the anti-symmetric 
fluxons with even topological charges is exactly
zero while for the odd charges it equals $4\eta$. The difference
comes from the Josephson energy of the horizontal phases at the 
$n=n_0$ site where $\phi^{(h)}_{n_0,1}=-\phi^{(h)}_{n_0,2}=Q\pi$. 
Each of these junctions contributes the value of $2\eta$ to the total 
energy. When $Q$ is even these sites correspond to the energy minima,
and, consequently, the total fluxon energy is zero. 
The energy of the asymmetric states is always zero (up to 
a constant $D_Q$). The energy of the multi-fluxon state 
defined by Eqs. (\ref{vert2})-(\ref{hor2a}) is a sum of the energies
of the constituting elementary fluxons. Here lies the difference between
the elementary fluxon (\ref{vert1})-(\ref{antsym1}) with some
even topological charge and the
multi-fluxon state (\ref{vert2})-(\ref{hor2a}) with two fluxons 
with odd topological charges $Q_1$ and $Q_2$, $Q_1+Q_2=Q$. Provided 
$\gamma=0$, the energy of the
former state equals $0$  while the energy of the latter
state is $4\eta (Q_1+Q_2)$. In the analogous case of asymmetric elementary and
multi-fluxon states the energy would be zero (if $\gamma=0$) in both 
cases. 
The difference would be hidden in the constant $D_Q$ which should
be redefined appropriately for the multi-fluxon case.

\begin{figure}[htb]
\includegraphics[scale=0.38]{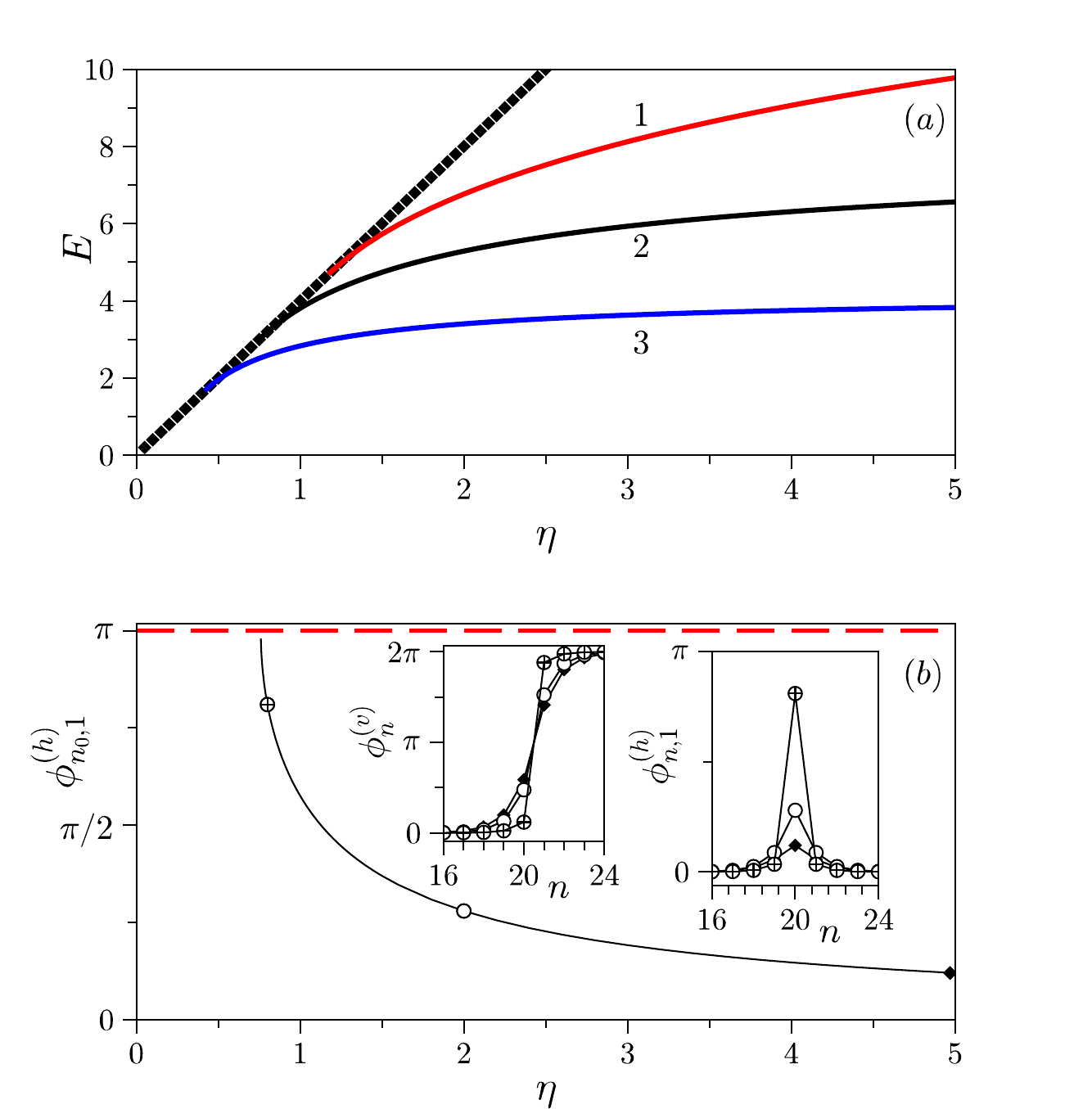}
\caption{(a) Energy of the non-compact stationary fluxons as
a function of the anisotropy $\eta$ for $\beta_L=0.25$ (line 1, red), 
$\beta_L=1$ (line 2, black) 
and $\beta_L=3$ (line 3, blue). The straight line ($\blacklozenge$) corresponds
to the energy of the compact fluxon, $E=4\eta$. \\
(b) Maximal value of the horizontal bottom phase, $\phi^{(h)}_{n_0,1}$
as a function of anisotropy for $\beta_L=1$. The dashed red line corresponds to the value $\phi^{(h)}_{n_0,1}=\pi$. The insets demonstrate the vertical $\phi^{(v)}_{n}$ and
bottom horizontal $\phi^{(h)}_{n,1}$ distributions for 
$\eta=5$ ($\blacklozenge$), $\eta=2$ ($\circ$) and  $\eta=0.8$ ($\oplus$).
The markers on the curve correspond to the above metioned values
of $\eta$.}
\label{fig5}
\end{figure}

In order to compare the energies of the non-compact and compact
fluxon states the total energy (\ref{ham2}) was computed numerically
as a function of the anisotropy $\eta$ and for the different values
of the discreteness constant $\beta_L$. The results for the
unbiased ($\gamma=0$) $Q=1$ fluxon are shown in Fig. \ref{fig5}(a). 
The non-compact fluxon state is computed by integrating numerically
Eqs. (\ref{emd1})-(\ref{emd7}) with the initial conditions (\ref{ic1}).
We have started from the limit of large $\eta$. In this case the
system converges to the non-compact stationary fluxon. The additional
initial distortion (\ref{ic2}) of the horizontal subsystem is not 
important, the system settles on the same stationary state.
As the 
anisotropy is decreased the fluxon energy decreases as well. At some
critical anisotropy, $\eta_c$, the fluxon energy equals
the respective energy of the anti-symmetric compact state, $4\eta$.
At this point and further on, for $\eta<\eta_c$, the system settles
on the anti-symmetric {\it compact} fluxon state.   
%
%
%
The point $\eta=0$ should be excluded, because we have division by
zero in Eqs. (\ref{emd2}). However, in this case the horizontal system
is completely decoupled from the vertical subsystem which is governed
by the standard DSG equation.

The non-compact fluxon energy is larger for the smaller values
of $\beta_L$ because the inductive energy is proportional to
$1/\beta_L$. Also, the critical anisotropy, $\eta_c$, increases when
$\beta_L$ is decreased. It is instructive to see how the non-compact
fluxon shape changes with the decrease of $\eta$. In Fig. \ref{fig5}(b)
we show how the maximum of the bottom horizontal phase 
distribution changes with $\eta$. The insets depict the fluxon profiles
in the vertical and bottom horizontal subsystems. We observe that
the fluxon becomes more localized as $\eta$ is decreased. 
The top horizontal phase is not plotted because $\phi^{(h)}_{n_0,2}=-\phi^{(h)}_{n_0,1}$. The maximal value $\phi^{(h)}_{n_0,1}$
also increases when $\eta$ is decreased until it reaches the
value of $\pi$ at $\eta=\eta_c$. Below $\eta<\eta_c$ only compact
fluxons exist.

It is of interest to see what happens if one simulates the 
equations of motion with compact initial conditions. We have
found that states with zero energy, i.e., the anti-symmetric with
even topological charges and asymmetric ones, persist. 
The compact anti-symmetric odd-charged fluxon initial conditions evolve
into the non-compact fluxon for $\eta>\eta_c$ and remain 
compact for $\eta<\eta_c$. However, these states are not
fully stable either. We have found that they are unstable
against only one type of perturbation. The  
deviation of the central horizontal phase, either 
 $\phi^{(h)}_{n_0,1}=\pi+\epsilon$ or $\phi^{(h)}_{n_0,2}=-\pi+\epsilon$,
facilitates the transition to the asymmetric state (\ref{asym1})
or (\ref{asym2}). Any other perturbation, including the 
anti-symmetric one of the type $\phi^{(h)}_{n_0,1}=\pi+\epsilon$, 
$\phi^{(h)}_{n_0,2}=-\pi-\epsilon$, leaves the anti-symmetric
odd-charged fluxon intact. The possible explanation must
be connected with flat band in the linear wave spectrum \cite{mffzp01pre} and the consequent absence of dispersion.

\begin{figure}[ht]
\includegraphics[scale=0.37]{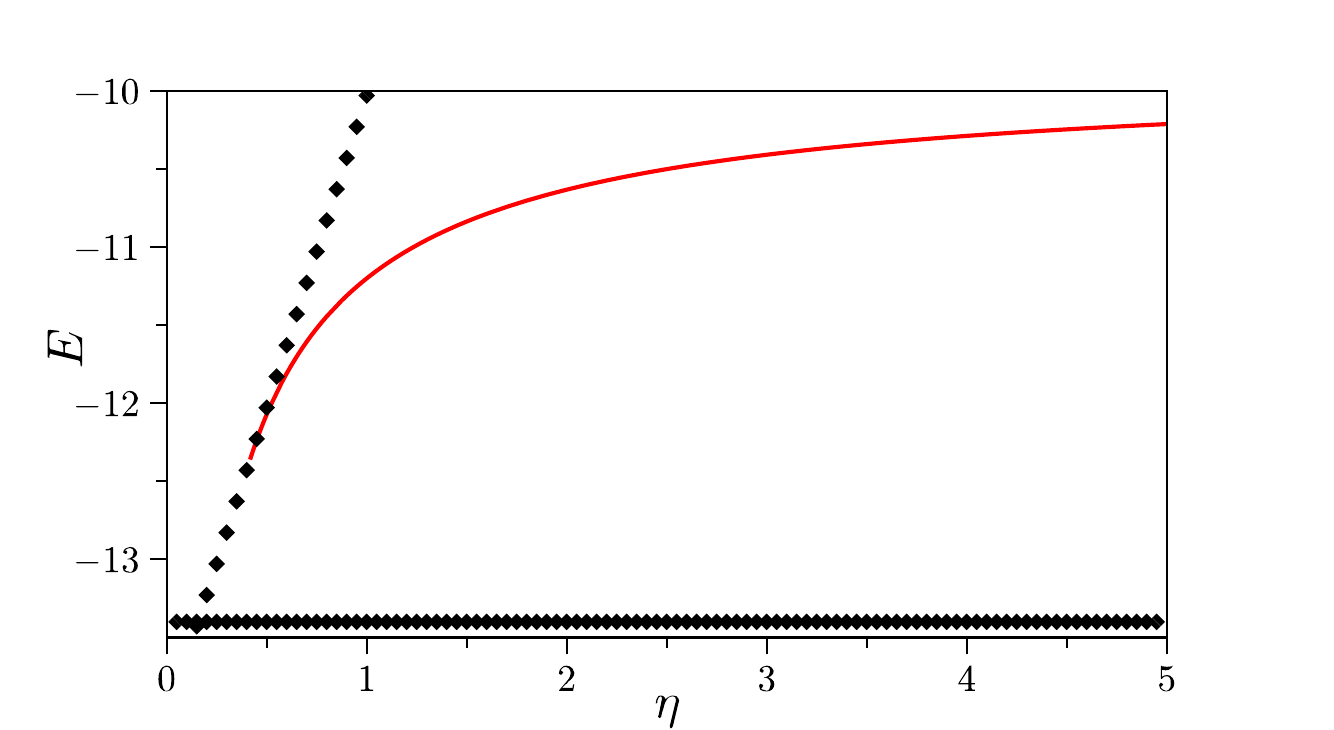}
\caption{Energy of the non-compact stationary fluxons as
a function of the anisotropy $\eta$ for $\beta_L=1$ and $\gamma=0.1$. 
The straight lines ($\blacklozenge$) corresponds
to the energy of the compact anti-symmetric and asymmetric fluxons.  }
\label{fig6}
\end{figure}

If the dc bias is applied, the general situation remains the same.
In Fig. \ref{fig6} we show again the energy of the non-compact fluxon
as a function of $\eta$ but for the biased JJL with $\gamma=0.1$.
%
%
The only difference with the unbiased case [Fig. \ref{fig5}(a)]
is in the absolute energy values since the constant $D_Q$ which is
given by Eq. (\ref{D}) is no longer zero.

\begin{figure}[ht]
\includegraphics[scale=0.37]{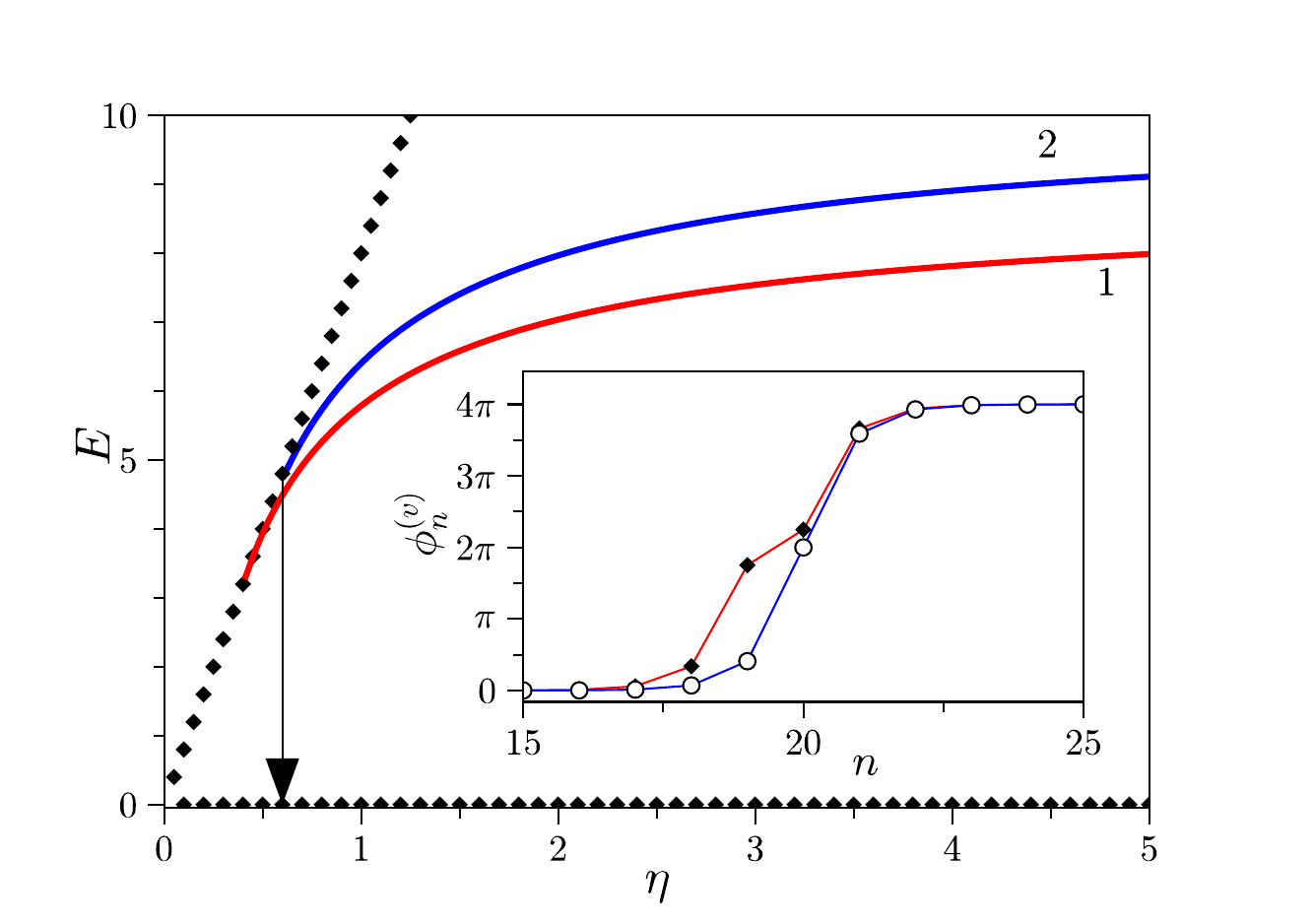}
\caption{Energy of the non-compact cell-centered (line 1, red) 
and junction-centered (line 2, blue) stationary fluxons with topological
charge $Q=2$ as
a function of the anisotropy $\eta$ for $\beta_L=3$ and $\gamma=0$. 
The straight lines ($\blacklozenge$) corresponds
to the energy of the compact anti-symmetric ($E=8\eta$) and asymmetric ($E=0$) bi-fluxons. The inset shows the cell-centered ($\blacklozenge$) and junction-centered $\circ$ fluxon profiles at $\eta=2$. }
\label{fig7}
\end{figure}

Now we turn our attention towards the multi-fluxon stationary states. 
We repeat the same numerical experiment by starting with the non-compact
bi-fluxon ($Q=2$) state and reducing the anisotropy. 
The multi-fluxon stationary states  demonstrate more complex behavior.
Firstly, the multi-fluxon family is much diverse because there is
an additional possibility to have different spatial separation between
the individual fluxons. Also, both the {\it stable} junction-centered (with
the fluxon center on a vertical junction) and cell-centered  (the
fluxon center lies in the middle of the junction cell, or, alternatively,
in the center of the junction cell) can exist. We have found 
only stable cell-centered $Q=1$ fluxons [see the left inset
in Fig. \ref{fig5}(b)]. In Fig. \ref{fig7} the dependence of the
non-compact fluxon energy on the anisotropy is shown. 
In order to obtain a multi-fluxon with a certain symmetry
we have used different parameter $\delta$ in the initial 
condition (\ref{ic1}). We have observed in our simulations that 
for $\delta=0,0.1,0.2$ the system is attracted to the junction-centered
state and for $\delta=0.4,0.5$ is is attracted to the cell-centered
state. The cell-centered 
bi-fluxon has smaller energy and transforms into the compact
anti-symmetric $(\ldots 0,2\pi,2\pi,4\pi,\ldots )$ bi-fluxon at 
$\eta\approx 0.5$ when its energy coincides with the $8\eta$ line. The
junction-centered non-compact bi-fluxon (red line) has larger energy  
and jumps into the compact elementary asymmetric $Q=2$ fluxon (\ref{vert1})-(\ref{asym1}) with zero energy. The critical anisotropy value 
$\eta_c$ that separates the compact and non-compact states is larger
for the bi-fluxon with the junction-centered symmetry.
%
%

We have also performed simulations for the smaller values of $\beta_L$.
Smaller $\beta_L$ causes the non-compact fluxons to be more
spread out. As a result, the critical point $\eta=\eta_c$ the
compactification process brings a composite bi-fluxon of the
type (\ref{vert1})-(\ref{antsym1}) with larger
separation between the constituent fluxons. Compactification of the
$Q=3$ multi-fluxons was also observed.

\begin{figure}[ht]
\includegraphics[scale=0.37]{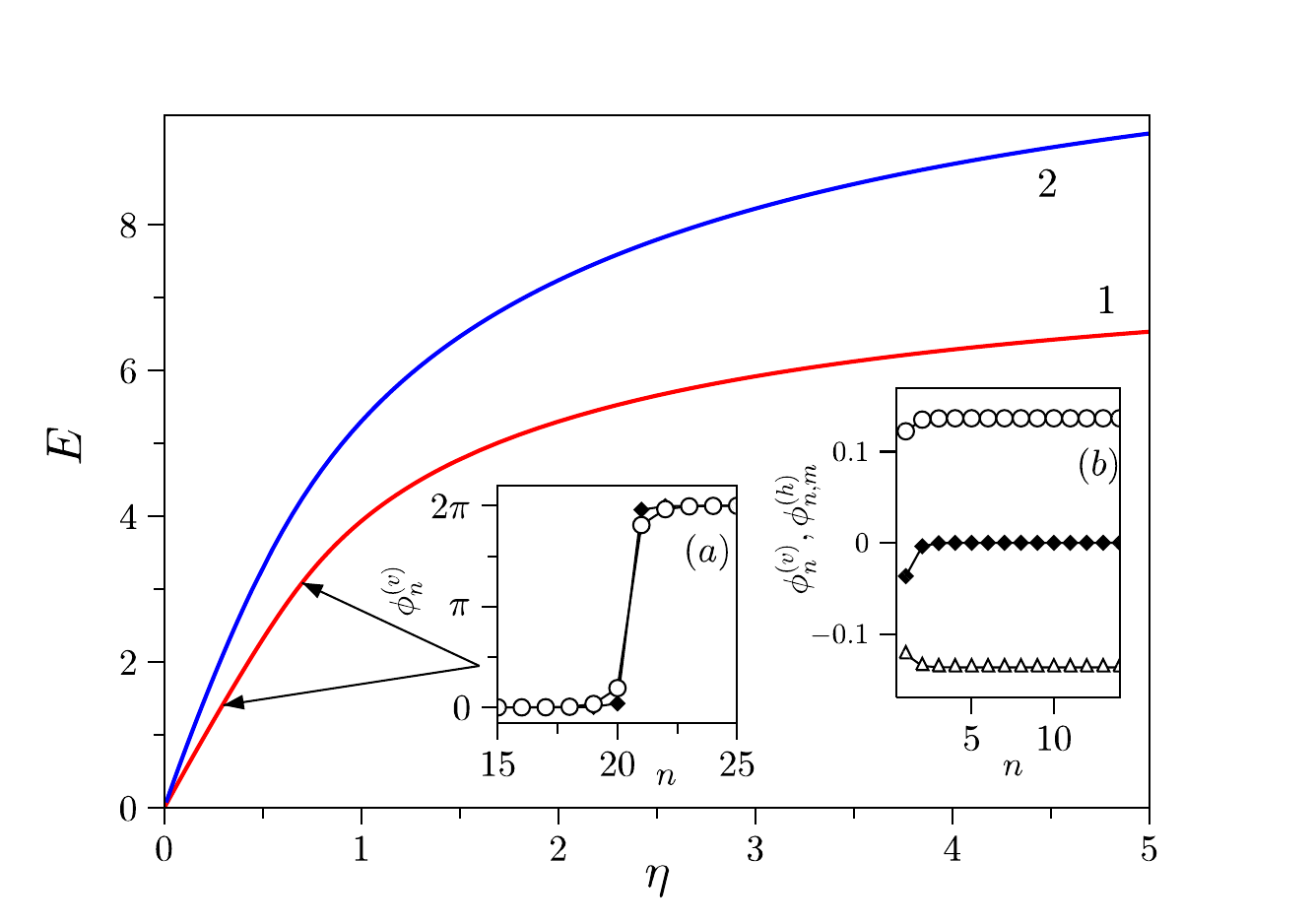}
\caption{Stationary fluxon energy
a function of $\eta$ for $\beta_L=1$, $\gamma=0$ and
magnetic field $f=0.05$ (line 1, red) and $f=0.1$ (line 2, blue). 
The inset (a) shows the central part of the 
vertical phase distribution for $\eta=0.3$ ($\blacklozenge$)
and $\eta=0.7$ ($\circ$). The inset (b) depicts the tail behavior
of the three components, $\phi_{n}^{(v)}$ ($\blacklozenge$), 
$\phi_{n,1}^{(h)}$ ($\triangle$) and $\phi_{n,2}^{(h)}$ ($\circ$)
for $\eta=0.3$, $f=0.05$.
 }
\label{fig8}
\end{figure}

The application of magnetic field ($f\neq 0$) destroys the compact solutions since
it becomes impossible to satisfy the stationary part of Eqs. (\ref{emd1})-(\ref{emd7}) in such a way that all the phases are the same.  
However, if the magnetic field is small, the non-compact states
are strongly localized as the anisotropy approaches zero. The
energy-anisotropy dependencies for the elementary fluxons is shown in
Fig. \ref{fig8}. The functions $E(\eta)$ are smooth, but their
behavior at small $\eta$ is almost linear. 
%
%
The profiles of the vertical phase distribution [see the inset (a) in Fig.{\ref{fig8}] show that although
the fluxons are no longer exactly compact, they are very strongly
localized. The fluxon tails show small distortion cause by the breaking
of the translation symmetry by the field.

\section{Compact fluxon interaction with the plane waves} \label{dyn}

Finally we investigate the interaction of stationary compact
fluxons with the small-amplitude waves of the ladder (Josephson
plasmons). In order to launch a plasmon we replace the
dc bias $\gamma$ in the equation for the $n=1$ vertical
junction by the mixed dc-ac bias $\gamma+\gamma_{ac}\sin \omega t$.
The stationary compact anti-symmetric fluxon with $Q=1$ 
is placed in the middle of the ladder
at $n_0=N/2$. The results of simulations of equations of motion
(\ref{emd1})-(\ref{emd7}) are shown in Fig. \ref{fig9}. The dissipation constant was chosen
rather small ($\alpha=0.01$) in order to minimize the spatial
and temporal decay of the plasmon
amplitude. The total integration time was $t=2 \times 10^4$.
So, the three-component plasmon wave is scattered on the stationary
compact fluxon. The plasmon dispersion law has three branches \cite{mffzp01pre} has three branches, one is strongly dispersive,
one is weakly dispersive and one is flat with $\Omega=1$. The
weakly dispersive branch is also flat if $\gamma=0$. For the
JJL parameters $\eta=0.5$, $\beta_L=1$ the frequencies of the
strongly dispersive branch lie in the interval $\sqrt{5}<\Omega<3$.
The highest plasmon density of states is achieved on the borders
of the plasmon band \cite{bsz25pla}. Therefore, we have chosen two
values of the driving frequency, one close to the bottom of the
band, $\omega=2.3$, and one close to the top of the band,
$\omega=2.98$.
%
%
%
\begin{figure}[ht]
\includegraphics[scale=0.37]{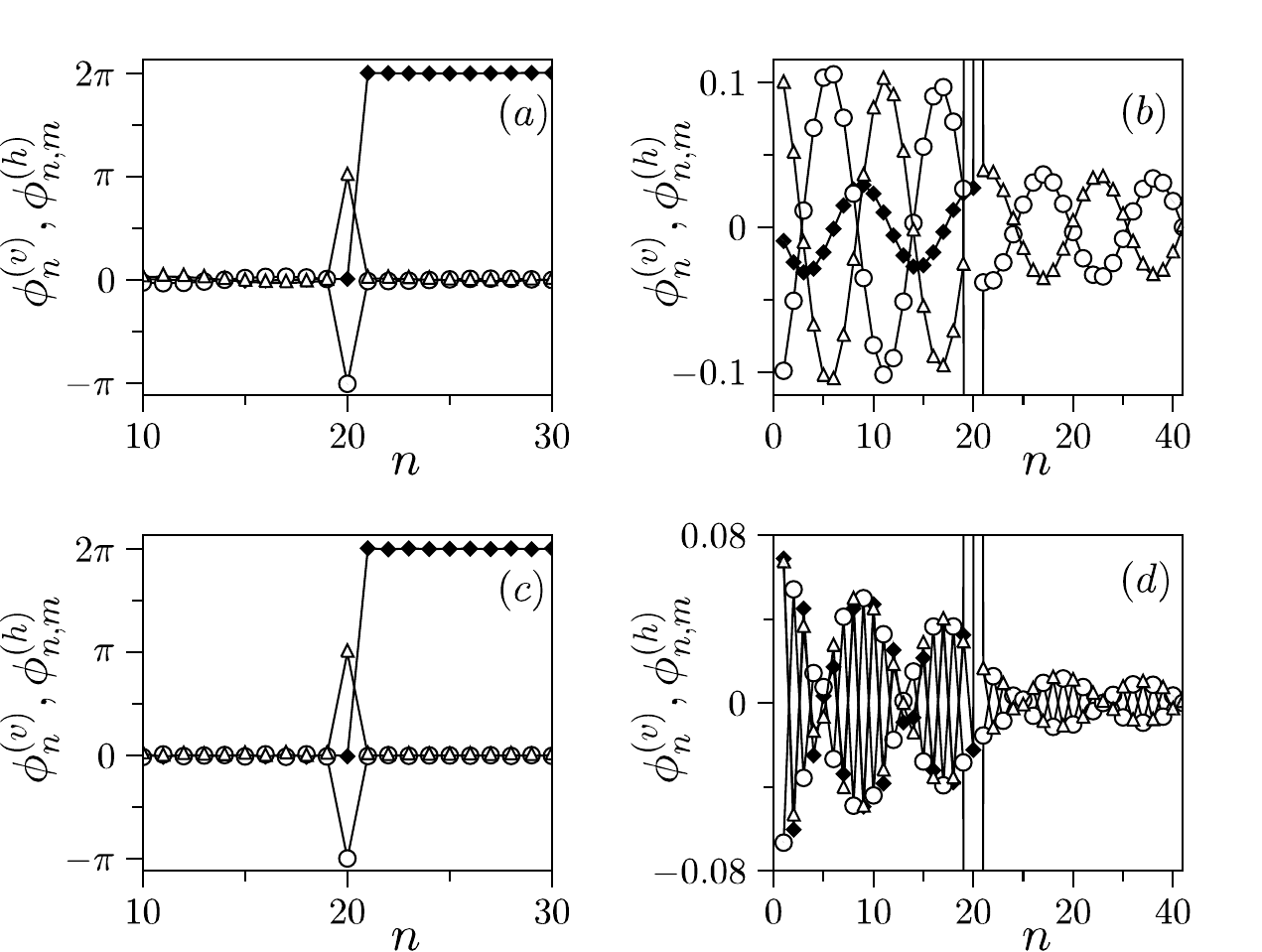}
\caption{Final fluxon profiles $\phi_{n}^{(v)}$ ($\blacklozenge$), 
$\phi_{n,1}^{(h)}$ ($\triangle$) and $\phi_{n,2}^{(h)}$ ($\circ$)
in the ladder with $N=40$, $\eta=0.5$, $\beta_L=1$ and $\alpha=0.01$.
The ac bias parameters are $\gamma=0$, $\gamma_{ac}=0.3$, $\omega=2.3$ [(a) and (b)]
and $\omega=2.98$ [(c) and (d)]. The panels (b) and (d) show the
details of the plasmon distribution for the panels (a) and (c), 
respectively.}
\label{fig9}
\end{figure}
As a result of interaction the compact fluxon survives and preserves
its strongly localized shape [see Fig. \ref{fig9}(a) and \ref{fig9}(c)]. 
The plasmon transmission is not perfect as the outgoing plasmon
amplitude is approximately halved. 
Part of the plasmon energy is absorbed by the fluxon. 
Earlier in this paper we have stated that the $Q=1$ anti-symmetric
compact fluxon is not stable. However, the plasmon is polarized in
such a way that the top-bottom antisymmetry is preserved. This
is clearly seen in Figs. \ref{fig9}(b) and \ref{fig9}(d). Thus, the
fluxon receives the anti-symmetric perturbation, and, as it was stated
before, it is stable against such a perturbation. More detailed research
on the plansmon-fluxon interaction is underway and will be reported
elsewhere.

\section{Conclusions} \label{conc}

In this work a new type of topological states that exist
in Josephson junction ladders (JJLs) has been discovered. These states 
are stationary compact fluxons that connect different ground states of the ladder and are exact solutions of the respective equations of motion. 
They are compact in the strict sense: the respective 
Josephson phases attain the values that are {\it exact} equilibrium
values of the individual junctions throughout the whole ladder.
In other words, the phase distribution in the subsystem of the
vertical junctions is an exact step-like function of the
junction number. The energies of the compact states can be written explicitly.
They are stable for the
 wide domain of the system parameters. They coexist
with the non-compact solutions if the anisotropy constant $\eta$ is large enough
and are the only stable stationary states if $0<\eta<\eta_c(\beta_L)$.
This can be interpreted as an integrable sector of the JJL, since the only
existing topological states are the compact ones and their energy is known.

Compact stationary fluxons exist for the dc-biased ladder, however,
their compactness is destroyed by any (even very small) external magnetic
field. The existence of compact states is connected to the
fact that the linear dispersion law of the JJL has a flat band. 
Hence, apart from the states that decay exponentially [${\phi_n} (n\to -\infty) \sim e^{-\lambda n}$] and are
non-compact, there can be states whose decay factor $\lambda$ is
exactly zero. Compact states are not possible in the parallel Josephson
junction arrays. The parallel array is governed by the DSG equation 
and supports only exponentially decaying non-compact fluxons.

\section*{Acknowledgements}
We would like to thank the Armed Forces of Ukraine for providing security
to perform this work.
I.O.S. and Y.Z. acknowledge support from the National
Research Foundation of Ukraine, grant (2023.03/0097)
"Electronic and transport properties of Dirac materials and
Josephson junctions".


\end{document}